\iftrue
\documentclass[aps,prl,twocolumn,
			   groupedaddress,superscriptaddress,
			   amsfonts,amssymb,amsmath,
			   citeautoscript,
			   a4paper]{revtex4-1}
\else
\documentclass[aps,pra,preprint,
			   groupedaddress,superscriptaddress,notitlepage,
			   amsfonts,amssymb,amsmath,
			   citeautoscript,
			   a4paper]{revtex4-1}
\fi

\usepackage[pdftex]{hyperref}
\hypersetup{colorlinks,
			linkcolor={blue!75!black!80!yellow},
			citecolor={blue!75!black!80!yellow},
			urlcolor={blue!75!black!80!yellow},
			pdfstartview=FitH}
\usepackage{graphicx}
\usepackage{mathrsfs}
\usepackage{amsthm}
\usepackage{physics}
\usepackage{xspace}
\usepackage{braket}
\usepackage{xr}
\usepackage{gensymb} 
\usepackage{xcolor,soul}
\usepackage{stmaryrd} 
\usepackage[UKenglish]{babel}
\usepackage{placeins} 
\usepackage{siunitx}
\DeclareSIUnit[number-unit-product=]\percent{\char`\%} 
\usepackage{makecell}
\usepackage{diagbox}
\usepackage{float}
\usepackage{amsmath} 
\usepackage{empheq} 

\usepackage{txfonts}  
\usepackage{txfontsb} 

\renewcommand{\Re}{\operatorname{Re}}
\renewcommand{\Im}{\operatorname{Im}}
\newcommand{\iu}{\mathrm{i}}
\newcommand{\e}{\mathrm{e}}

\newcommand{\appropto}{\mathrel{\vcenter{
			\offinterlineskip\halign{\hfil$##$\cr
				\propto\cr\noalign{\kern2pt}\sim\cr\noalign{\kern-2pt}}}}}

\newcommand{\ie}{i.e.\@\xspace}  
\newcommand{\cf}{cf.\@\xspace}
\newcommand{\eg}{e.g.\@\xspace}

\newcommand{\BdG}{\text{BdG}} 
\newcommand{\PBC}{\text{PBC}} 
\newcommand{\OBC}{\text{OBC}} 

\makeatletter
\newcommand*{\addFileDependency}[1]{
  \typeout{(#1)}
  \@addtofilelist{#1}
  \IfFileExists{#1}{}{\typeout{No file #1.}}
}
\makeatother

\usepackage{scalerel}

\usepackage{textcomp} 
\usepackage{xifthen}
\usepackage{etoolbox}
\newboolean{togglecomments}
\newboolean{togglechanges} 

\setboolean{togglecomments}{false}
\setboolean{togglechanges}{true}

\newcommand{\comment}[2]{%
    \ifbool{togglecomments}%
    {\textcolor{blue!70!black}{\small\textsf{%
    \textsuperscript{\textsc{\textsf{\MakeLowercase{#1}}}}%
    [#2]}}} 
    {}}     
\newcommand{\swap}[2]{\ifbool{togglechanges}
    {#2}  
    {\textcolor{red!70!black}{[#1]}\textrightarrow{}\textcolor{green!50!black}{[#2]}}}
\newcommand{\remove}[1]{\ifbool{togglechanges}
    {}    
    {\textcolor{red!70!black}{#1}}}
\newcommand{\inset}[1]{\ifbool{togglechanges}
    {#1}  
    {\textcolor{green!50!black}{#1}}}
\newcommand{\optional}[1]{\ifbool{togglechanges}
    {#1}  
    {\textcolor{yellow!50!orange!80!gray}{#1}}}
\newcommand{\citeremind}[1]{%
    [\textcolor{blue!75!black!80!yellow}{
        $\blacksquare$%
           \ifthenelse{\isempty{#1}}
               {}
               {\textsuperscript{\textsf{#1}}}%
        }]\xspace}
\newcommand{\todo}[1]{
    \textcolor{orange!80!yellow!95!black}{\textbf{[}%
        \ifthenelse{\isempty{#1}}%
        {\text{$\blacksquare$}}%
        {{\small\textsf{#1}}}%
        \textbf{]}}}
        
\newcommand{\hkuaffil}{\footnotesize Department of Physics and HK Institute of Quantum Science and Technology, The University of Hong Kong, Pokfulam, Hong Kong, China}
\newcommand{\hkustate}{\footnotesize State Key Laboratory of Optical Quantum Materials, The University of Hong Kong, Pokfulam, Hong Kong, China}

\begin{document}

\preprint{APS/123-QED}

\title{Self-partitioned Interfacial Time Crystals}

\author{Joseph Huang}%
\email{hqnj@connect.hku.hk}
\author{Yi Yang}
\email{yiyg@hku.hk}
\affiliation{\hkuaffil}
\affiliation{\hkustate}

\date{\today}

\begin{abstract}
Nonequilibrium many-body systems can spontaneously break symmetry in time, as in time crystals, or in space, through self-organized domains and interfaces. Whether these two forms of symmetry breaking can intertwine so that an emergent interface alone hosts time-crystalline order remains unknown.
In this work, by introducing the Rabi-Hatano-Nelson model, we unveil the existence and mechanism of a self-partitioned interfacial time crystal (SPITC), where a homogeneous system generates its own internal and tunable interface, at which the time-translation symmetry is also spontaneously broken.
Such a SPITC phase is intrinsically induced by nonreciprocity and open boundary conditions, without external pumping or long-range interaction.
The periodic and open boundary phase diagrams of the system are both mapped out; vacuum and Dicke-like superradiance with static or active orders are identified, with analytical phase boundaries in the weak coupling limit.  
The frequency of the SPITC is found to scale quadratically with the spin-photon coupling strength, as we derive analytically for the slow dynamics of the spins.
The position of the SPITC boundary scales with a critical exponent of $-1$ as a function of the degree of nonreciprocity, in stark contrast to $-1/2$ for an otherwise stationary boundary.
Our construction of SPITC establishes a route to spatiotemporal order in non-Hermitian many-body systems.

\end{abstract}

\maketitle

Nonreciprocity—the lack of symmetry between forward and backward processes—provides a route to nonequilibrium phases with no equilibrium analogue, generating directional transport, synchronization, boundary accumulation, and traveling collective states \cite{Hatano1996Localization, fruchart2021non,you2020nonreciprocity,Hanai2024Nonreciprocal, Khasseh2025Active, Avni2025Nonreciprocal, veenstra2025nonreciprocal, Jana2026Quantum}. In open systems, these effects are especially pronounced because gain, loss, and reservoir engineering can convert directionality into genuinely many-body dynamical order, including traveling-wave states, synchronization transitions, and boundary-sensitive critical behavior \cite{Zhang2022Symmetry, Nadolny2025Nonreciprocal, belyansky2025phase, Begg2024Quantum, Many2024Skin}. Recent progress in chiral quantum optics and waveguide QED shows that asymmetric light--matter couplings are scalable and can support many-body photon dynamics, making interacting photonic lattices a natural platform for studying strongly correlated photons \cite{lodahl2017chiral, suarez2025chiral, Wanjura2020Topological, Sheremet2023Waveguide, Tang2022Nonreciprocal, Mandal2020Nonreciprocal}.

In particular, strongly interacting photonic systems have opened a route toward realizing quantum many-body physics with light \cite{Carusotto2013Quantum, Noh2016Quantum, Hartmann2016Quantum}. The essential mechanism is that strong light-matter coupling converts otherwise non-interacting photons into nonlinear polaritonic excitations, enabling photon blockade \cite{mahmoodian2020dynamics, ridolfo2012photon}, effective photon-photon interactions \cite{calajo2022emergence, douglas2016photon, tevcer2024strongly}, and collective phases analogous to those of condensed-matter lattice models \cite{Greentree2006Quantum}. Foundational proposals showed that arrays of coupled cavity-QED units can realize Mott-insulator-to-superfluid physics of light, formalized in the Jaynes-Cummings (JC)-Hubbard model \cite{koch2009superfluid, Hartmann2006Strongly, angelakis2007photon, mahmoodian2019chiral, caleffi2023collective}, while later work established cavity and circuit-QED lattices as a broader platform for quantum simulation with interacting photons \cite{Houck2012On}. Going beyond the rotating-wave approximation, the Rabi-Hubbard model incorporates counter-rotating processes and therefore captures regimes of ultrastrong coupling, broken excitation-number conservation, and qualitatively different normal-superradiant phase transition from its JC counterpart \cite{beaudoin2011dissipation, Schiro2012Phase, Schiro2016Exotic, cui2020nonlinear, ye2021quantum}. These models have become paradigmatic because they connect quantum optics, strongly correlated matter, and nonequilibrium physics within a single framework, and recent experiments have begun to access them directly, including trapped-ion realizations of the Rabi-Hubbard model \cite{Mei2022Experimental} and many-body simulations of JC-Hubbard dynamics \cite{Li2022Observation}. Against this backdrop, introducing nonreciprocity and dissipation into interacting photonic lattices is a natural and timely step toward uncovering genuinely new phases of correlated light in which strong interactions, openness, and boundary-sensitive non-Hermitian effects act on equal footing \cite{Chiacchio2023Nonreciprocal}.

\begin{figure}
\includegraphics[width=0.8\linewidth]{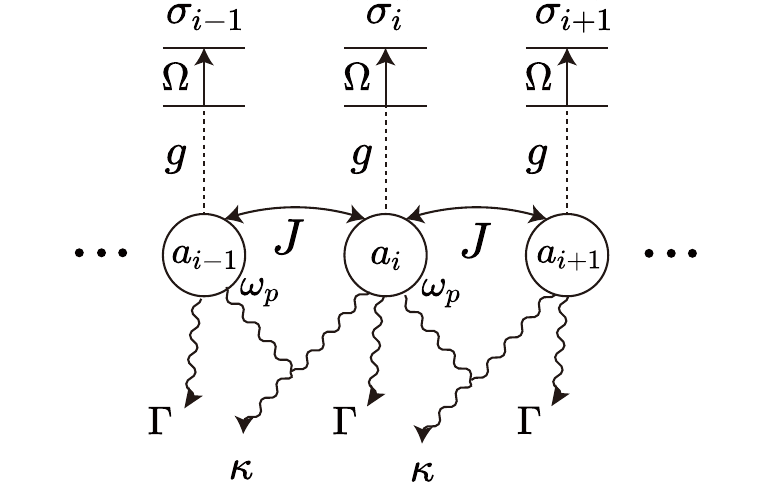}
\caption{\label{fig:model}
\textbf{Model sketch.} The system is a one-dimensional bosonic chain, and each site couples to a two-level system without conserving excitation number. The bosons are dissipative with single-site uncorrelated loss $\Gamma$ and nearest-neighbor correlated loss $\kappa$.
}

\end{figure}

Such interplay of nonreciprocity, dissipation, and interactions naturally provides a fertile land for time-dependent ordered phases. Time crystals---phases characterized by persistent oscillations that break continuous or discrete time---translation symmetry—have been extensively studied \cite{wilczek2012quantum, Shapere2012Classical, Bruno2013Impossibility, Watanabe2015Absence, Heugel2019Classical, kongkhambut2022observation, Zaletel2023Colloquium}. 
Since the original proposal of time crystals, the notion of a boundary time crystal (BTC) was originally introduced to denote a phase in which time-translation symmetry breaking occurs only in a macroscopic fraction of the system \cite{iemini2018boundary}; however, most existing realizations that are labeled as BTC effectively describe collective oscillations of the entire open system viewed as the “boundary” of an external bath, as in long-range or collective-spin models where the order parameter is global of the system of interest rather than spatially localized \cite{iemini2018boundary, wang2025boundary, carollo2022exact}. Beyond this collective scenario, related forms of spatially inhomogeneous temporal order have also been explored, including chimera time-crystalline phases in which distinct dynamical orders coexist in different regions \cite{sakurai2021chimera, belyansky2025phase}, and nonlinear skin breathing modes in nonreciprocal or non-Hermitian topological systems, where time-periodic motion is pinned near a system boundary \cite{manda2026nonlinear,lang2021non}. 
Nevertheless, all existing time crystals break time translational symmetry either globally or in a pre-defined subset of a
system; it is thus tempting to ask: whether an initially homogeneous open many-body system can self-organize simultaneously in both space and time? 

We answer this question affirmatively by discovering the self-partitioned interfacial time crystal (SPITC), a qualitatively different form of boundary temporal order: the oscillation is neither a global collective mode of the whole system, nor a coexistence of externally partitioned time-crystalline orders, nor a system boundary-pinned nonlinear breather inherited from a conventional skin mechanism; instead, it involves a spontaneously formed, dynamically stabilized interface between distinct phases, which do not have time crystalline order themselves, and the position and internal coherence of the interface exhibit robust, long-lived oscillations. This distinction becomes evident considering the fact that the “interface” here is spontaneously formed as an actual spatial subset of the system rather than the full open system itself.
This work thus establishes a new mechanism for boundary time crystallinity rooted in the interplay of nonreciprocity and strong interactions, and highlights the fundamentally different role of boundaries in non-Hermitian many-body systems compared to both equilibrium lattices and previously studied dissipative time-crystalline phases.

\textit{Model.} Our model, which we term the Rabi-Hatano-Nelson model, consisted of three parts, the onsite light-matter interaction $H^{\text{Rabi}}_{j}$, photon hopping between nearest-neighbor sites $H^{\text{ph}}_{j} =-J(a^\dagger_j a_{j+1} + a^\dagger_{j+1} a_{j})$ and single-particle photon loss. The overall Hamiltonian is $H=\sum_j H^{\text{Rabi}}_{j} + H^{\text{ph}}_{j}$, where
\begin{align}
    H^{\text{Rabi}}_{j} = (\Omega/2) \sigma^z_j + \omega_0 a_j^\dagger a_j + g \sigma^x_j (a^\dagger_j + a_j).
\end{align}
The parameter $\Omega$ denotes the level splitting of the two-level system, $\omega_0$ is the cavity photon frequency, and $g$ characterizes the light–matter coupling strength. The interaction term includes both energy-conserving and counter-rotating processes and it is important in the nonequilibrium dynamics.
This open system is described by the Lindblad master equation $\partial_t \rho = -\iu [H,\rho] + \sum_{j,n} D[L^{n}_j]\rho$. The jump operator for correlated loss $L^{1}_j=\sqrt{2\kappa}(a_j+\iu\e^{\iu\theta} a_{j+1})$ and uncorrelated loss $L^{2}_j=\sqrt{2\Gamma}a_j$ with dissipators $D[L_j]=L_j \rho L_j^\dagger - \frac{1}{2}\left\{L_j^\dagger L_j , \rho\right\}$ describe the incoherent single particle decay rates $\Gamma$ and $\kappa$. In the limit of decoupled matter and light part, the Heisenberg equation of motion of $a_j$ is governed by the dissipative Hatano-Nelson model $H_{\text{NH}} = \sum_j \left(\omega_0-\iu (\Gamma+2\kappa)\right) a^\dagger_j a_j - \left(J+\kappa\e^{-\iu\theta}\right) a_{j}^\dagger a_{j+1} -\left(J-\kappa\e^{\iu\theta}\right) a^\dagger_{j+1} a_j$ in the form of $\iu\partial_t a_j = \sum_i (H_{\text{NH}})_{ji}a_i$.
In this work, we study the mean-field dynamics of the master equation,
\begin{equation}
\begin{aligned}
    \partial_t \alpha_j & = -(\Gamma+2\kappa+\iu\omega_0) \alpha_j - \iu g s^x_j - \iu J_+\alpha_{j-1} - \iu J_-\alpha_{j+1},
    \\
    \partial_t \vec{s}_j & = \vec{s}_j \times \left(2g(\alpha_j + \alpha_j^*),0,\Omega \right)^{\mathrm{T}},
\end{aligned}
\label{eq:ODE}
\end{equation}
where $\alpha\equiv\langle a\rangle$ and $\vec{s}=\langle\vec{\sigma}\rangle$ are the expectation values, and we have defined $J_+ \equiv -(J+\kappa \e^{-\iu\theta})$ and $J_- \equiv -(J-\kappa \e^{\iu\theta})$.

\begin{figure}
\includegraphics[width=\linewidth]{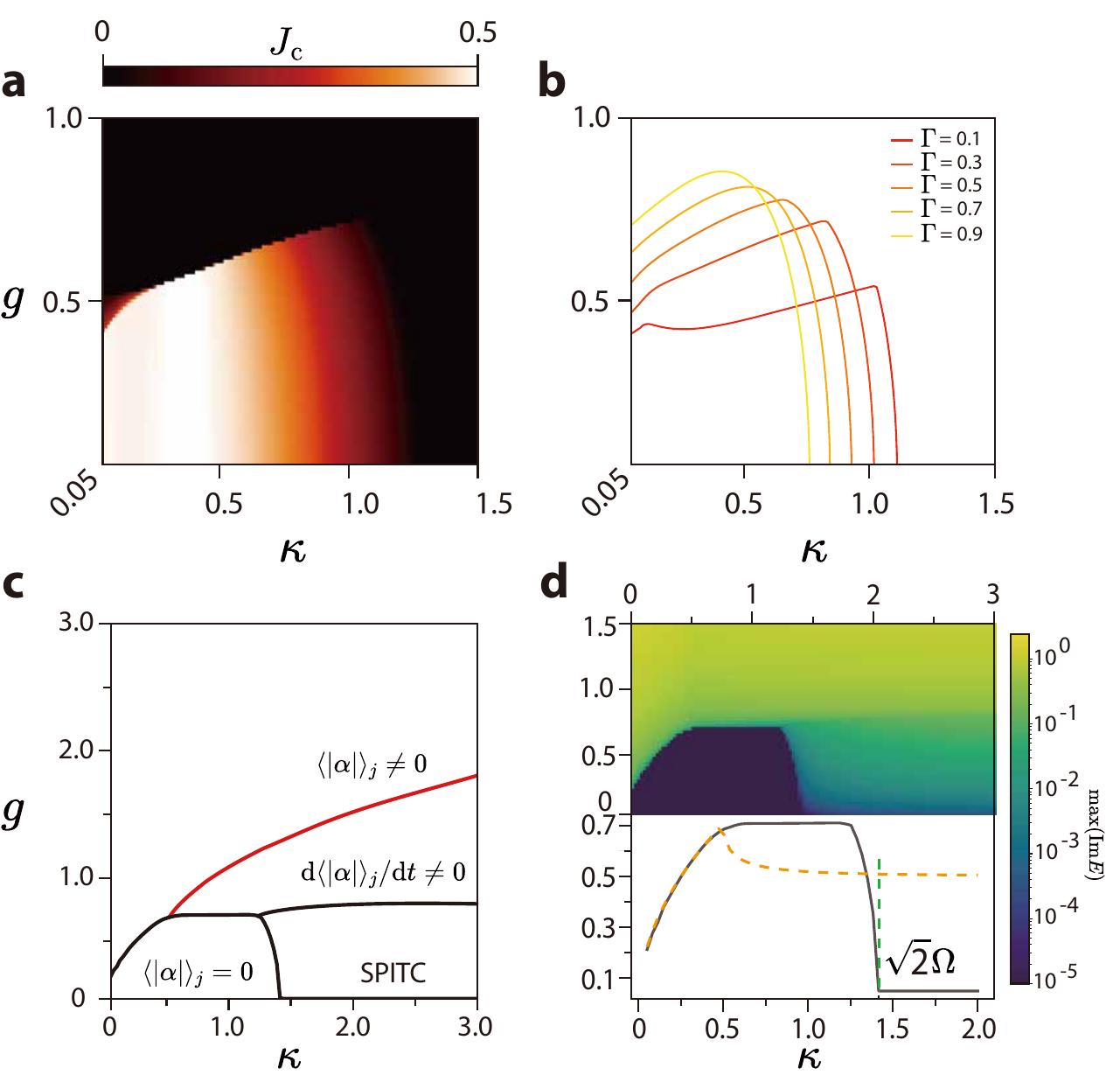}
\caption{\label{fig:Phase}
    \textbf{PBC and OBC Phase diagrams of the Rabi-Hatano-Nelson model}. 
    \textbf{a.} The heatmap of vacuum-superradiant PBC phase boundary shown via the critical $J_\mathrm{c}^\PBC(\kappa,g)$. $\omega_0=\Omega=1$, $N=400$ and $\Gamma=0.5$. 
    \textbf{b.} The contour map of $J_\mathrm{c}^\PBC=0.2$ PBC phase boundary for different $\Gamma$. 
    \textbf{c.} The OBC phase diagram. Black lines are well described by the BdG Hamiltonian, while red lines are extrapolated from the late-time behavior of Eq.~\eqref{eq:ODE}. 
    \textbf{d.} Upper panel: $\max\mathrm{Im} E$ of $H_\BdG$ under OBC. Lower panel: Analytical understanding of the boundary [Eq.~\eqref{eq:AnalyticalOBCBoundary} (orange dashed line) and Eq.~\eqref{eq:smallJkappa} (green vertical dashed line) compared with the numerical diagonalization (black line) of the real space BdG Hamiltonian with $N=40$ sites. 
    In all figures here, $\omega_0=\Omega=1$ and $J=0.5$.
}

\end{figure}

\textit{Periodic Boundary Condition (PBC).}
We now turn to the discussion of PBC, where the pure photon part of the Hamiltonian can be diagonalized in momentum space with complex spectrum $\omega_q=\omega_0 - 2J\cos(q) -\iu\Gamma- 2\iu\kappa(1+\sin(q-\theta))$. We focus on maximal nonreciprocal settings $\theta=0$ or $\pi$ below (See SM S1A for more general situations). The instability onset of the vacuum ($\alpha_j=0$, $\vec{s}=(0,0,-1)$) is analyzed from the spectrum of the Jacobian of Eq.~\eqref{eq:ODE}.

In the absence of single-site loss, \ie $\Gamma=0$, the PBC system is always unstable for all $g$, $\kappa$, and $J\geq0$, under which the system contains one lossless mode $q*$ where nonreciprocity and dissipation compensate each other (see SM S1C). The phase boundary $J_\mathrm{c}^\PBC(g,\kappa)$ associated with the critical hopping strength that separates the stationary vacuum and superradiant phase can be solved from the equations (see SM S1A),
\begin{equation}
    \begin{aligned}
    & \left(\Omega^2-\omega^2 \right) \left(\left(\Gamma+2\kappa \right)^2 - 4\kappa^2\sin^2(q) + \Delta_q^2 - \omega^2 \right) = 4g^2\Omega\Delta_q,
    \\
    & \left(\Omega^2-\omega^2 \right) \left(\left(\Gamma+2\kappa \right)\omega + 2\kappa\sin(q) \Delta_q \right) = 4g^2\Omega\kappa\sin(q),
\end{aligned}
\label{eq:PBCInstability}
\end{equation}
where $\omega$ is the Hopf bifurcation frequency of the system and we defined $\Delta_q\equiv\Delta(q)=\omega_0-2J\cos(q)$. 
For $\theta=0$ or $\pi$, the least dissipative mode locates at $q^*=\pm\pi/2$. 
Because $\Delta(\pm\pi/2)$ is independent of $J$, no nontrivial $J_\mathrm{c}^\PBC$ exists if the system destabilizes through the mode $q^*$. 
Plugging $q^*$ into Eq.~\eqref{eq:PBCInstability} solves a $J$-independent boundary $g_\mathrm{c}(\kappa)$ (analytically obtained in SM S1A). Shown in Fig.~\ref{fig:Phase}a, it corresponds to the boundary where $J_\mathrm{c}^\PBC$ jumps to zero at small $\kappa$ when $g$ gradually increases. We can further prove that it is almost always a Hopf instability $\omega\neq0$ if system destabilizes through the mode $q^*$ (SM S1A).  
One can solve Eq.~\eqref{eq:PBCInstability} for $(\omega, \Delta_q)$. Once $\Delta_q$ is known, the critical $J_\mathrm{c}^\PBC=\min_q J_\mathrm{c}^\PBC(q)$ can be obtained from the definition of $\Delta_q$. As nonreciprocity increases with $\kappa$, the first destabilized mode $q$ shifts to $q^*$ and $J_\mathrm{c}^\PBC$ gradually drops to zero, see bottom right corner of Fig.~\ref{fig:Phase}a.
Besides, the mode $q=0$ gives the solution of Eq.~\eqref{eq:PBCInstability} as $\mathrm{max}(J_\mathrm{c}^\PBC)=\omega_0/2$, imposing a universal upper bound of $J_\mathrm{c}$ independent of $g$, $\kappa$, and $\Omega$ (Fig.~\ref{fig:Phase}a).

\textit{Open Boundary Condition (OBC).} We now turn to the study of the OBC of the system where SPITC appears. Its phase diagram for small $J/\Omega$ (the case of large $J/\Omega$ is discussed below) is shown in Fig.~\ref{fig:Phase}c.
The phase diagram is drastically different from the case of PBC. Because the boundary is an effective loss channel, a vacuum phase exists even when $\Gamma=0$.
At small $\kappa$, the system behaves like the conventional Dicke model, where it is in the vacuum with zero photon field on every site $\alpha_j=0$ for small $g$.
The system goes into a superradiant state with nonzero photon distribution $\langle|\alpha|\rangle_j\neq0$ at large $g$. 
At intermediate $\kappa$ (\eg, $\kappa\approx1$), there are two phase transition points $g_\mathrm{c}$. The first one lies between the vacuum and a chaotic phase, where the fields are excited but also actively evolving with nonzero time derivative $\mathrm{d}\langle|\alpha|\rangle_j/\mathrm{d}t\neq0$. As $g$ further increases, the system passes the second phase boundary and enters the static superradiant phase. As $\kappa$ further increases, a new order which we identify as self-partitioned interfacial time crystal phase appears at small nonzero $g$ (bottom right corner of Fig.~\ref{fig:Phase}c).

To understand this phase diagram, we perform the Holstein–Primakoff transformation such that $\sigma^x \approx b+b^\dagger$ where $b$ is a bosonic operators obeying the bosonic commutation relation $[b,b^\dagger]=1$. Defining $\Psi_q = (a_q, a^\dagger_{-q}, b_q, b^\dagger_{-q})$, the linearized equation of motion $\partial_t \Psi_q = -\iu H_\BdG(q)\Psi_q$ is governed by the Bogoliubov de-Gennes (BdG) Hamiltonian. Interestingly, the BdG method successfully captures all the phase boundaries at small g (black lines in Fig.~\ref{fig:Phase}c). 
It is also worth mentioning that the PBC BdG Hamiltonian relates to the Jacobian of the vacuum of Eq.~\eqref{eq:ODE} via a similarity transformation, and, therefore, they exhibit the same spectrum (see SM S1B).
\begin{figure*}
\includegraphics[width=\linewidth]{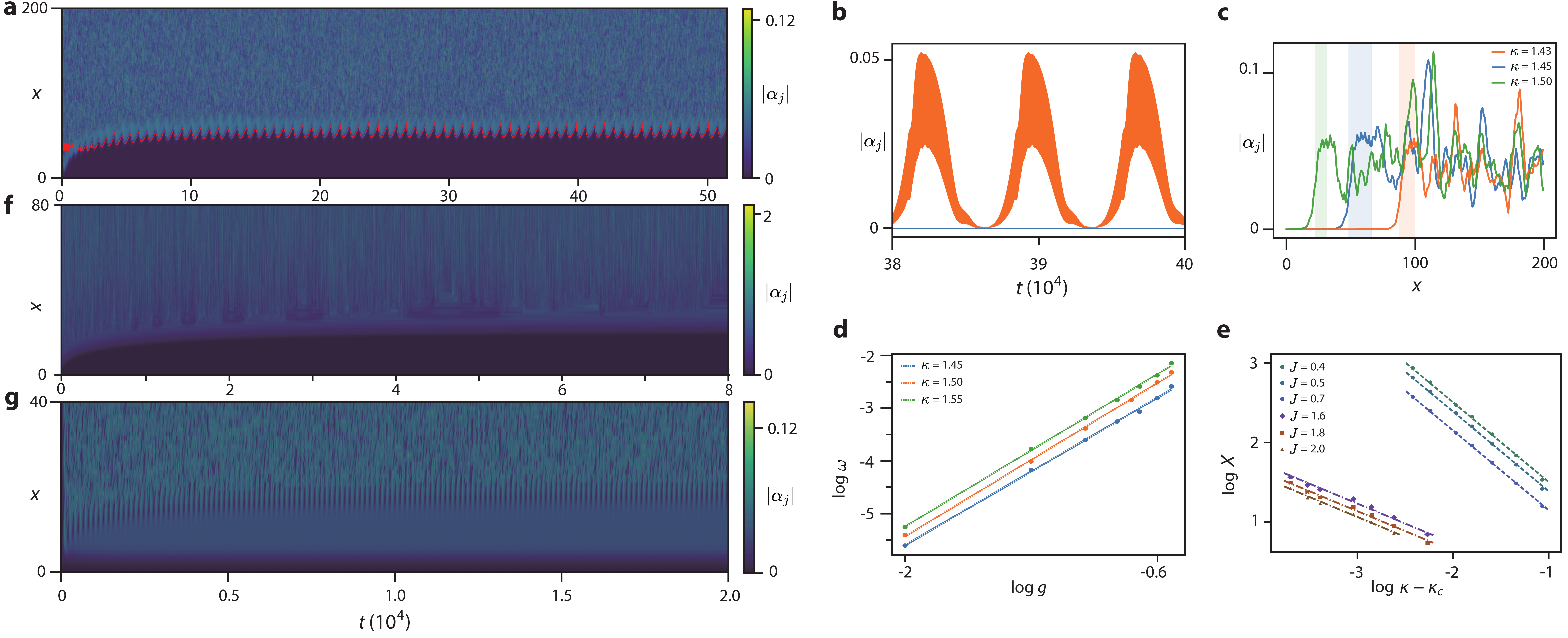}
\caption{\label{fig:SPITC}
\textbf{Self-partitioned interfacial time crystal.} 
    \textbf{a.} The space-time diagram for a $\Gamma=0$ Rabi-Hatano-Nelson model when time evolves from a random initial state. Here $J=0.5$, $g=0.2$, $\kappa=1.45$ and $\omega=\Omega=1$. See SM 2D for the rule of drawing the red curve.
    \textbf{b.} Dynamics of $|\alpha_j|$ in (a) on the 20th site (blue line) and the 50th site (orange line), located outside and within the oscillatory front, respectively. 
    \textbf{c.} Time snapshots of the smoothed photon field in (a) at time $t\approx4.0\times10^{5}$ for different $\kappa=1.43,1.45,1.5$.
    The shaded regions indicate the location of the spontaneously formed interfacial oscillation. 
    \textbf{d.} The scaling of SPITC frequency as a function of light-matter coupling strength $g$ near the critical point $g_{\text{crit}}=0$ at different $\kappa=1.45, 1.50, 1.55$, with fitted slopes $2.001, 2.081, 2.068$, respectively. 
    \textbf{e.} The scaling of the front position $X$ as a function of nonreciprocity $\kappa$ near the critical point $\kappa_\mathrm{c}$, which is $-1$ under $J/\Omega=0.4,0.5,0.7<\sqrt{3/2}$ for the oscillatory front and $-1/2$ at $J/\Omega=1.6,1.8,2.0>\sqrt{3/2}$  for the stationary front, respectively. 
    \textbf{f.} The space-time diagram for a stationary front. Here $J=1.5$, $g=1.0$, $\kappa=1.533$ and $\omega=\Omega=1$.
    \textbf{g.} The space-time diagram of a more complicated form of SPITC at the interface between the chaos and superradiance regions when the single particle loss rate is nonzero $\Gamma = 1.6$. Here $J= 0.5$, $g = 0.6$, $\kappa= 0.7$ and $\omega= \Omega = 1$.
}

\end{figure*}

The upper panel of Fig.~\ref{fig:Phase}d shows the largest imaginary part $\max(\Im E_n)$ among $4N$ complex energies on a log scale. 
Evidently, $\max(\Im E)=0$ in the vacuum phase region, whereas $\max(\Im E)$ exhibits an order-of-magnitude difference between the superradiance region and SPITC region. 
We analytically obtain part of the vacuum-superradiance phase boundary (orange dashed line in the bottom panel of Fig.~\ref{fig:Phase}d) via
\begin{align}
    \left(g_\mathrm{c}(\kappa)\right)^2 = \frac{\Omega}{4} \min_m \left\{
         \mathcal{E}(k_m)+ \frac{\kappa^2}{\mathcal{E}(k_m)} 
        \right\},
    \label{eq:AnalyticalOBCBoundary}
\end{align}
for $\kappa<J$, where $k_m=m\pi/(N+1)$, $m=1, \dots,N$, and $\mathcal{E}(k_m)=\omega_0 - 2 \sqrt{J_+ J_-} \cos(k_m)$. It is obtained through imaginary gauge transforming the real space BdG Hamiltonian to reproduce reciprocal hopping between sites (see SM S2A).

This method of imaginary gauge transformation fails at large $\kappa$ region, nonetheless it can be overcome by separating the slow dynamics $u$ and fast precession $\e^{\iu \Omega t}$ of the spins such that $s^+\equiv s^x+\iu s^y\equiv u(t) \e^{\iu\Omega t}$, where in this region the local photon dynamics is always fast and respond to the fast spin precession (see SM 2B). As a result, the photon field can be solved from spins. At late time,
\begin{align}
    \alpha_j(t) & = - \frac{\iu g}{2} \left( [R_+ u]_j \e^{\iu \Omega t} + [R_- u]^*_j \e^{-\iu \Omega t} \right),
    \\
    R_\pm & \equiv \left(\pm\iu\Omega I + \iu H_{\text{HN}} \right)^{-1}.
\end{align}
The photon field strength $|\alpha_j|$ thus depends linearly on $g$. With known photon fields in terms of spins, we can write down the effective equations that are in closed form of spins (see SM 2B),
\begin{align}
    \frac{\partial}{\partial T} u_j & = s_j^z
    [Du]_j,
    \\
    \frac{\partial}{\partial T} s^z_j & = -\Re\{u_j^*[Du]_j \},
    \label{eq:effective equation}
\end{align}
where we defined the matrix $D$ and slow time coordinate $T$ as
\begin{align}
    D & = R_-^* - R_+, \\
    T & = g^2 t.
\end{align}
This set of equations captures the slow envelope of the system. Its linearized version $\partial_T u = -Du$ indicates that $D$ encodes the core information about the system, including its instability onset when the largest real part of eigenvalues of $D$ reaches zero, \ie $\max\Re D(k)=0$, in which the first destabilized mode $k^*$ satisfies $\partial_k \Re D(k^*)=0$. The linearized equation now permits the method of imaginary gauge transformation in extended $\kappa$ region. At zero detuning $\omega_0=\Omega$ (see SM S1C for more general situations),
\begin{align}
    \kappa_\mathrm{c} = \sqrt{2} \Omega, \quad\;\;\qquad \text{if} \;\; J<J_\mathrm{c}^\OBC\equiv\sqrt{\frac{3}{2}}\Omega,
    \label{eq:smallJkappa}
\end{align}
which is shown as the blue dashed line in Fig.~\ref{fig:Phase}(d). $\kappa_c$ is interestingly independent of $J$ when $J<J_\mathrm{c}^\OBC$. Higher-order corrections give $g$-dependent terms that appear in the exact phase boundary in Fig.~\ref{fig:Phase}(d).

\emph{Self-partitioned interfacial time crystals.}
For small $g$, the system enters the SPITC phase when $\kappa>\kappa_\mathrm{c}$. In this phase, there is a robust coexistence: one side of the chain remains in a vacuum-like state while the rest of the system develops an active and chaotic bulk (Fig.~\ref{fig:SPITC}a). The two regions are separated by a sharp front whose mean position depends strongly on \(\kappa\) but only negligibly on \(g\) in the small-\(g\) regime. Meanwhile, the front does not remain stationary: it undergoes a regular oscillation with a characteristic frequency that scales as \(\omega_{\rm front}\propto g^2\) for fixed \(\Omega\), \cf Eq.~\eqref{eq:effective equation}.

The robustness of this oscillatory front is examined in the thermodynamic limit. As the system size $N$ is increased, the size of the vacuum region remains approximately constant on one side, whereas the additional sites are all fall into the chaotic domain, as long as the model parameters are fixed. Also, arbitrary random initial states fall into the SPITC attractor in late time, where the system self-organizes without imposing any predefined front structure. This initial state is chosen to be isotropic and homogeneous, such that there is no artificial grouping of order or disorder initially.
The resulting space-time heatmap is shown in Fig.~\ref{fig:SPITC}(a), and the dynamics of single sites in the SPITC region in a late time window is plotted in Fig.~\ref{fig:SPITC}(b).
The location of the oscillatory front can be seen from 
the field distribution for different $\kappa$, as shown in Fig.~\ref{fig:SPITC}(c). Note that the amplitude of the front plateau is nearly constant when $g$ remains fixed. 
Additionally, we find strong numerical evidence that the SPITC frequency $\omega_{\text{SPITC}}$ scales near the critical point as $(g-g_{\text{crit}})^{\nu}$ with $\nu\approx2$ and $g_{\text{crit}}=0$, as shown in Fig.~\ref{fig:SPITC}(d).

These observations admit a natural interpretation in terms of a nonreciprocal boundary instability problem of the vacuum as an effective smaller system of size $X$, which can be solved by the perturbation analysis of the total system (SM S2C).
In the SPITC phase, there is a nonvanishing coefficient $\partial_k d(k^*)$ of the first-order contribution of $k$ in $\Re D(k)=0$, and it is purely imaginary. Its existence is a direct consequence of $2\Omega$ phase lag between the two resolvents $R_-$ and $R_+$ in $D$. Furthermore, the self-organized boundary allows a complex $k^*_{\text{eff}}$ rather than a real $k$ on a physical boundary (\eg $k^*$). These two imaginary factors hence give a nonzero first order contribution of $1/X$ in the critical stability condition of the effective smaller system $\Re D(k^*_{\text{eff}})=0$ (since $k\sim1/X$). This explains the existence of SPITC: a self-organized front with phase gradient can absorb or leak. As another immediate result, it gives the scaling law
\begin{align}
    X \sim (\kappa - \kappa_\mathrm{c})^{-1}.
\end{align}

On the other hand, when $J>J_\mathrm{c}^\OBC$, there are two critical points on the $\kappa$ axis (SM S2C and Fig.~S2):
\begin{align}
    \kappa_{\mathrm{c}1} = \sqrt{J^2-\frac{\Omega^2}{4}},
    \quad
    \kappa_{\mathrm{c}2} = \frac{2J^2 - \Omega^2}{2\sqrt{J^2-\Omega^2}},
    \quad \text{if} \; J>J_\mathrm{\mathrm{c}}^\OBC.
    \label{eq:BigJkappa}
\end{align}
Vacuum phase appears when $\kappa_{\mathrm{c}1}<\kappa<\kappa_{\mathrm{c}2}$ (SM S2C and Fig.~S2).
In contrast, when either condition, $\kappa<\kappa_{\mathrm{c}1}$ or $\kappa>\kappa_{\mathrm{c}2}$, is satisfied, the system can exhibit a stationary front (Fig.~\ref{fig:SPITC}f) instead of SPITC. 
This is because the first destabilized mode $k$ falls on the band edge, \ie $k=0$ or $\pi$, and by symmetry $\partial_k d(k^*)=0$ vanishes. The leading contribution is second order in $1/X$, and it gives the scaling law 
\begin{align}
    X \sim (\kappa - \kappa_\mathrm{c})^{-\frac{1}{2}}.
\end{align}
This difference between the critical exponents clearly distinguishes SPITC and a stationary interface (Fig.~\ref{fig:SPITC}e).

When we turn on the single-site dissipation ($\Gamma>0$), both critical $\kappa_\mathrm{c}$ and $J_\mathrm{c}$ [Eq.~\eqref{eq:smallJkappa} and Eq.~\eqref{eq:BigJkappa}] gets modified (see SM S2C). Furthermore, there can appear a more complicated form of SPITC (Fig.~\ref{fig:SPITC}g). In this case, there are three different regions in the system: vacuum, uniform oscillation, and chaotic. SPITC exists on the layer between the chaotic region and the uniform oscillation region, and the SPITC involves a slow and a fast oscillation frequency, where the fast frequency equals the frequency of the uniform oscillation region.

\textit{Conclusions and outlook.} In summary, we show the existence of a self-partitioned interfacial time crystal and study it within a dissipative nonreciprocal extension of the Rabi-Hubbard model, termed the Rabi-Hatano-Nelson model.
By providing effective equations and analytical phase boundary formulae, we show that nonreciprocity fundamentally reshapes both its phase diagrams and long-time dynamics under different boundary conditions. Under open boundary conditions, the boundary acts as an effective dissipation channel, stabilizing a vacuum phase even without on-site loss and enabling SPITCs. This SPITC frequency scales as $\omega_\text{SPITC} \sim g^2$ and its front position scales as $(\kappa-\kappa_\mathrm{c})^{-1}$ as a key property of SPITCs. In comparison, we also show that the exponent reduces to $1/2$ for a stationary front. 

A realistic route to implement the present model is a one-dimensional superconducting circuit-QED lattice composed of microwave resonators, each locally coupled to a superconducting qubit. In the near-resonant and rotating-wave regime, this architecture realizes Jaynes-Cummings-Hubbard physics, while in the ultrastrong-coupling regime, or in a parametrically synthesized rotating frame, it can be extended toward an effective Rabi-Hubbard description. The nonreciprocal dissipative hopping can be engineered by combining bare coherent photon tunneling between neighboring resonators with reservoir-mediated dissipative coupling through auxiliary lossy modes or waveguide segments \cite{metelmann2015nonreciprocal, wang2024dispersive, bernier2017nonreciprocal}. Since large-scale dissipative circuit-QED lattices \cite{fitzpatrick2017observation}, ultrastrong-coupling \cite{baust2016ultrastrong, bosman2017multi} or analog-synthesized Rabi interactions \cite{braumuller2017analog} and reservoir-engineered nonreciprocity \cite{bernier2017nonreciprocal} have all been demonstrated separately in superconducting or closely related photonic platforms, their integration into a driven-dissipative circuit-QED array could serve as a promising platform for realizing the phenomena predicted in this work.


\bibliographystyle{apsrev4-2}
\bibliography{sources}

\end{document}


\title{
Supplementary Materials\\Self-partitioned Interfacial Time Crystals}

\author{Joseph Huang}
\email{hqnj@connect.hku.hk}
\author{Yi~Yang}
\email{yiyg@hku.hk}
\affiliation{\hkuaffil}
\affiliation{\hkustate}

\maketitle

\setlength{\parindent}{0em}
\setlength{\parskip}{.5em}

\noindent{\textbf{\textsf{CONTENTS}}}\\ 
\twocolumngrid
\begingroup 
    \let\bfseries\relax 
    \deactivateaddvspace 
    \tableofcontents
\endgroup
\onecolumngrid

\hyphenation{eigen-index}

\section{Periodic Boundary Condition}

\subsection{Linear stability analysis}

We begin from the mean-field equations in order to identify when the spatially uniform vacuum first becomes unstable. Here, $\alpha_j$ is the coherent photon amplitude at site $j$, while $s_j^{x,y,z}$ are the components of the local Bloch vector. The parameters $\omega_0$ and $\Omega$ set the photon and two-level-system frequencies, $g$ is the onsite light--matter coupling, and $J_\pm$ describe directional photon hopping. The term $\Gamma+2\kappa$ gives the local decay scale associated with the effective non-Hermitian dynamics. The equations read
\begin{equation}
    \begin{aligned}
    \partial_t \alpha_j & = -(\Gamma+2\kappa+\iu\omega_0) \alpha_j - \iu g s^x_j - \iu J_+\alpha_{j-1} - \iu J_-\alpha_{j+1},
    \\
    \partial_t s^x_j & = -\Omega s^y_j,
    \\
    \partial_t s^y_j & = \Omega s^x_j - 2g(\alpha_j+\alpha_j^*) s^z_j,
    \\
    \partial_t s^z_j & = 2g(\alpha_j+\alpha_j^*) s^y_j.
\end{aligned}
\label{eq:SM_PDE}
\end{equation}

The normal state contains no coherent photons and has every spin polarized along down $s^z$ in its lower state. It is therefore represented by
\begin{align}
    \alpha_j=0, \quad s^x_j=0, \quad s^y_j=0, \quad s^z=-1,
\end{align}
To test its stability, we add small fluctuations in all dynamical variables,
\begin{align}
    \alpha_j=\delta\alpha_j, \quad s^x_j=\delta s^x_j, \quad s^y_j=\delta s^y_j, \quad s^z_j=-1+\delta s^z_j.
\end{align}
The equation for $\delta s^z_j$ contains products of $\delta \alpha$ and $\delta s^y$ (see Eq.~\eqref{eq:SM_PDE}), so its time dependence starts only at the second order. It consequently decouples from the remaining variables at the linear order. Dropping all quadratic and higher-order contributions gives
\begin{align}
    \partial_t \alpha_j & = -(\Gamma+2\kappa+\iu\omega_0) \alpha_j - \iu g s^x_j - \iu J_+\alpha_{j-1} - \iu J_-\alpha_{j+1},
    \\
    \partial_t s^x_j & = -\Omega s^y_j,
    \\
    \partial_t s^y_j & = \Omega s^x_j + 2g(\alpha_j+\alpha_j^*),
\end{align}
where the symbol $\delta$ is suppressed from now on for readability. Under periodic boundary conditions, translation invariance makes momentum a good quantum number. We therefore seek modes with lattice momentum $q$ and complex growth exponent $\lambda$,
\begin{align}
    \alpha_j = \sum_q \alpha_q \e^{\lambda t+\iu q j}, \quad s_x=\sum_q X_q \e^{\lambda t + \iu q j}, \quad 
    s_y=\sum_q Y_q \e^{\lambda t + \iu q j}.
\end{align}
where $\alpha_q$, $X_q$ and $Y_q$ are the Fourier components, and $\Re\lambda>0$ signals an exponentially growing perturbation. The directional hopping contribution in a given momentum sector can be written as
\begin{align}
    J_+ \e^{-\iu q} + J_- \e^{\iu q} = 2J\cos q - 2\iu \kappa \sin(q+\theta).
\end{align}
This expression separates the usual cosine dispersion from the momentum-dependent dissipative contribution produced by nonreciprocity. Since the term $\alpha_j+\alpha_j^*$ mixes a photon fluctuation with its complex conjugate, the $q$ and $-q$ sectors must be treated together. The resulting four-dimensional eigenmode problem is
\begin{align}
    J_q\begin{pmatrix}
        \alpha_q
        \\
        \alpha_{-q}^*
        \\
        X_q
        \\
        Y_q
    \end{pmatrix}
    \equiv
    \begin{pmatrix}
        -\delta_+(q) -\iu\Delta(q) & 0 & -\iu g & 0
        \\
        0 & -\delta_-(q) +\iu\Delta(q) & \iu g & 0
        \\
        0 & 0 & 0 & -\Omega
        \\
        2g & 2g & \Omega & 0
    \end{pmatrix}
    \begin{pmatrix}
        \alpha_q
        \\
        \alpha_{-q}^*
        \\
        X_q
        \\
        Y_q
    \end{pmatrix}
    =
    \lambda\begin{pmatrix}
        \alpha_q
        \\
        \alpha_{-q}^*
        \\
        X_q
        \\
        Y_q
    \end{pmatrix},
\end{align}
where $J_q$ is the Jacobian matrix, $\Delta(q)$ is the effective photon detuning at momentum $q$, and $\delta_\pm(q)$ are the damping rates experienced by the two conjugate photon components:
\begin{align}
    \Delta(q)&=\omega_0-2J\cos(q)
    \\
    \delta_+(q)&=\Gamma+2\kappa-2\kappa\sin(q+\theta)
    \\
    \delta_-(q)&=\Gamma+2\kappa+2\kappa\sin(q-\theta)
\end{align}
At an instability threshold, the leading eigenvalue reaches the real axis; this eigenvalue can thus be parametrized as $\lambda=\iu\omega$ in the stable region.
Separating the characteristic equation into its real and imaginary parts yields the two phase-boundary conditions
\begin{align}
    \left(\Omega^2 -\omega^2\right)\left(\delta_+(q)\delta_-(q)+\Delta_q^2-\omega^2\right) &= 4g^2\Omega\Delta_q,
    \\
    \left(\Omega^2-\omega^2\right)\left[\omega(\delta_+(q)+\delta_-(q))-\Delta_q\left(\delta_+(q)-\delta_-(q)\right)\right]&=-2g^2\Omega\left(\delta_+(q)-\delta_-(q)\right)
\end{align}

The critical point is obtained by satisfying both conditions for at least one momentum and then selecting the mode that destabilizes first.

We now focus on the maximally nonreciprocal regime, $\theta=0$ or $\pi$. In this case the photon damping is smallest at $q=\pm\pi/2$. Moreover, $\Delta(\pm\pi/2)=\omega_0$ is independent of $J$, so this mode provides a direct test of whether the system is already unstable when $J=0$. Substituting $q=\pi/2$ gives
\begin{align}
    \left(\Omega^2 - \omega^2 \right) \left[\Gamma(\Gamma+4\kappa)+\omega_0^2-\omega^2 \right] & = 4g^2\Omega\omega_0
    \\
    \left(\Omega^2 - \omega^2 \right) \left[\Gamma(\Gamma+2\kappa)\omega+2\kappa\omega_0 \right] & = 4g^2\Omega\kappa
\end{align}
For a static instability, $\omega=0$, the two conditions reduce to
\begin{align}
    g^2=\Omega\omega_0/2, \quad \omega_0^2=\Gamma(\Gamma+4\kappa),
\end{align}
This requires two independent constraints and therefore occupies only a line in the $(\kappa,g)$ plane. Away from this fine-tuned line, the least-damped $q=\pi/2$ mode crosses the stability boundary at a nonzero frequency, giving a Hopf instability. Solving the two conditions for this oscillatory onset gives the candidate frequencies and critical coupling
\begin{align}
    \omega_\pm = \frac{1}{2\kappa}\left(-(\Gamma+2\kappa)\omega_0\pm\sqrt{\Gamma(\Gamma+4\kappa)\left(\omega_0^2+4\kappa^2 \right)} \right),
    \\
    g^2_c=\min_\pm\left\{ \frac{1}{4\Omega\kappa}\left(\Omega^2-\omega_\pm^2\right)\left[(\Gamma+2\kappa)\omega_\pm+2\kappa\omega_0 \right]\right\},
\end{align}
Only branches that give a real frequency and a positive physical value of $g_c^2$ are retained in $\min\{\cdot\}$. The system is unstable already at $J=0$ whenever $g>g_c(\kappa)$. If this inequality is not satisfied, the first unstable mode need not be the least-damped one, and the threshold must be determined over all momenta:
\begin{align}
    \left(\Omega^2-\omega^2 \right) \left(\left(\Gamma+2\kappa \right)^2 - 4\kappa^2\sin^2(q) + \Delta_q^2 - \omega^2 \right) & = 4g^2\Omega\Delta_q,
    \\
    \left(\Omega^2-\omega^2 \right) \left(\left(\Gamma+2\kappa \right)\omega + 2\kappa\sin(q) \Delta_q \right) & = 4g^2\Omega\kappa\sin(q).
\end{align}
For each $q$, these equations determine the threshold frequency and effective detuning. The corresponding hopping strength follows from the definition of $\Delta_q$, and the global threshold is the smallest admissible value over all momenta,
\begin{align}
    J_c = \min_q \frac{\omega_0-\Delta_q}{2\cos(q)}.
\end{align}
The $q=0$ sector provides a useful reference because the nonreciprocal damping asymmetry vanishes there. In this sector, the second boundary condition reduces to
\begin{align}
    \left(\Omega^2-\omega^2 \right)  \omega = 0,
\end{align}
or $\omega=\Omega$, which leads to
\begin{align}
    \Delta_q=0.
\end{align}
So for $q=0$,
\begin{align}
    J_c=\frac{\omega_0}{2},
\end{align}
Thus the uniform mode becomes resonant when the bottom of the photon band reaches zero detuning. This threshold is independent of $g$, $\Omega$, and $\kappa$, and consequently $\omega_0/2$ supplies an upper bound on the critical $J_c$ obtained after minimizing over all modes.

\subsection{Bogoliubov–de Gennes analysis}

The same stability problem can be formulated equivalently by Bogoliubov-de Gennes (\BdG) formalism. We use the Holstein--Primakoff transformation \cite{Holstein1940Field} to represent small spin fluctuations above the fully polarized normal state as bosonic excitations,
\begin{align}
    S_+ = \sqrt{2s}b^\dagger \sqrt{1-\frac{b^\dagger b}{2s}}, \qquad S_- = \sqrt{2s}\sqrt{1-\frac{b^\dagger b}{2s}} b, \qquad S_z = -s+b^\dagger b,
\end{align}
where $b$ is a bosonic operator. The square-root factors enforce the finite spin Hilbert space and reproduce the spin commutation relations when $[b,b^\dagger]=1$. For a two-level system, $s=1/2$. Close to the normal state the density of $b$ excitations is small, so the onsite Rabi Hamiltonian may be expanded in powers of $b$:
\begin{align}
    H^{\text{Rabi}}_{j} & = \Omega \left(-\frac{1}{2}+b^\dagger b \right) + \omega_0 a^\dagger a + g \left(a^\dagger + a\right) \left(\left(1-\frac{b^\dagger b}{2} \right) b + b^\dagger \left( 1 - \frac{b^\dagger b}{2} \right) + O(b^5) \right)
    \\
    & \approx \Omega \left(-\frac{1}{2}+b^\dagger b\right) + \omega_0 a^\dagger a + g \left(a^\dagger + a\right) \left(b+b^\dagger \right).
\end{align}
Keeping only quadratic terms gives the quadratic Hamiltonian; higher powers describe interactions between fluctuations and are irrelevant for the onset of instability. Combining this approximation with photon hopping and the effective non-Hermitian loss, and dropping the constant energy shift, gives
\begin{align}
    H_{\BdG} = \sum_j \left(\left(\omega_0-\iu(\Gamma+2\kappa)\right) a^\dagger_j a_j + \Omega b^\dagger_j b_j + g\left(a_j + a_j^\dagger\right)\left(b_j + b_j^\dagger \right) \right) - \left( \left(J+\kappa\e^{\iu \theta}\right) a_j^\dagger a_{j+1} + \left(J-\kappa\e^{\iu \theta}\right) a_{j+1}^\dagger a_{j} \right)
\end{align}

With periodic boundaries, the Fourier transform
\begin{align}
    a_j = \frac{1}{\sqrt{N}} \sum_{q} \e^{\iu q j} a_q, \qquad b_j = \frac{1}{\sqrt{N}} \sum_q \e^{\iu q j} b_q.
\end{align}
diagonalizes the spatial part of the photon sector, with dispersion to be
\begin{align}
    H_{\text{ph}} & = \sum_q \left(\omega_0 - \iu(2\kappa+\Gamma) - t_R \e^{\iu q} - t_L \e^{-\iu q} \right) a^\dagger_q a_q
    \\
    & = \sum_q \left(\omega_0 - 2J\cos(q) - \iu(2\kappa+\Gamma+2\kappa\sin(q-\theta)) \right) a^\dagger_q a_q
\end{align}
The real part of this dispersion controls the photon frequency, while its imaginary part gives the momentum-dependent damping. Because the Rabi coupling contains both number-conserving and anomalous terms, it is convenient to introduce 
\begin{align}
    \Psi_q = (a_q, a^\dagger_{-q}, b_q, b^\dagger_{-q})
\end{align}
which groups the coupled $q$ and $-q$ particle--hole components. The linearized Heisenberg equation then takes the form
\begin{align}
    \partial_t \Psi_q= -\iu H_\BdG(q) \Psi_q ,
\end{align}
with the non-Hermitian BdG matrix
\begin{align}
    H_\BdG (q)= 
    \begin{pmatrix}
        \omega_q & 0 & g & g
        \\ 
        0 & -\omega_{-q}^* & -g & -g
        \\
        g & g & \Omega  & 0
        \\
        -g & -g & 0 & -\Omega
    \end{pmatrix},
    \label{eq:PBCBdG}
\end{align}
where $\omega_q=\omega_0 - \iu(2\kappa+\Gamma) - t_R \e^{\iu q} - t_L \e^{-\iu q}$. The complex eigenvalues give the oscillation frequencies and decay or growth rates of the elementary fluctuations. Their characteristic polynomial is
\begin{align}
    \det (H_\BdG - \lambda) & = (\lambda - \omega_{q})(\lambda + \omega_{-q}^*) (\lambda^2 - \Omega^2) - 2\Omega g^2 \left(\omega_{q} + \omega_{-q}^* \right)
    \\
    & = \left(\lambda^2 + \iu 4 \kappa \lambda - \left(a^2 + 4\kappa^2 \cos^2 q \right) + \iu 4 \kappa a \sin q \right) (\lambda^2 - \Omega^2) - 2\Omega g^2 \left(2a-\iu 4 \kappa \sin q \right)
\end{align}

The first term in this polynomial describes the uncoupled photon and spin branches, while the second term proportional to $g^2$ produces their hybridization. To connect this operator formulation with the preceding mean-field calculation, define $x_q=\langle b_q+b_{-q}^\dagger\rangle$ and $y_q=\iu\langle b_q-b_{-q}^\dagger\rangle$. Then $\left(\alpha_q,\alpha_{-q}^*,x_q,y_q\right)=T\left(\alpha_q,\alpha_{-q}^*,\beta_q,\beta_{-q}^*\right)$, where $\beta=\langle b\rangle$ and
\begin{align}
    T = \begin{pmatrix}
        1 & 0 & 0 & 0
        \\ 
        0 & 1 & 0 & 0
        \\
        0 & 0 & 1 & 1
        \\
        0 & 0 & \iu & -\iu
    \end{pmatrix}.
\end{align}
The relation $T(-\iu H_\BdG(q))T^{-1}=J_q$ 
is a similarity transformation. It shows explicitly that the mean-field Jacobian and the quadratic BdG theory yield identical linear stability spectra.

\subsection{Dynamical behaviors}

When $\Gamma=0$, the vacuum is always unstable if $g>0$ and $\kappa>0$. It is found that the Hopf frequency must be $\omega=-\omega_0$ and instability starts at $g=0$. The phases are identified from the Lyapunov exponent and mode distribution in late-time dynamics. 

\begin{figure}
\includegraphics[width=\linewidth]{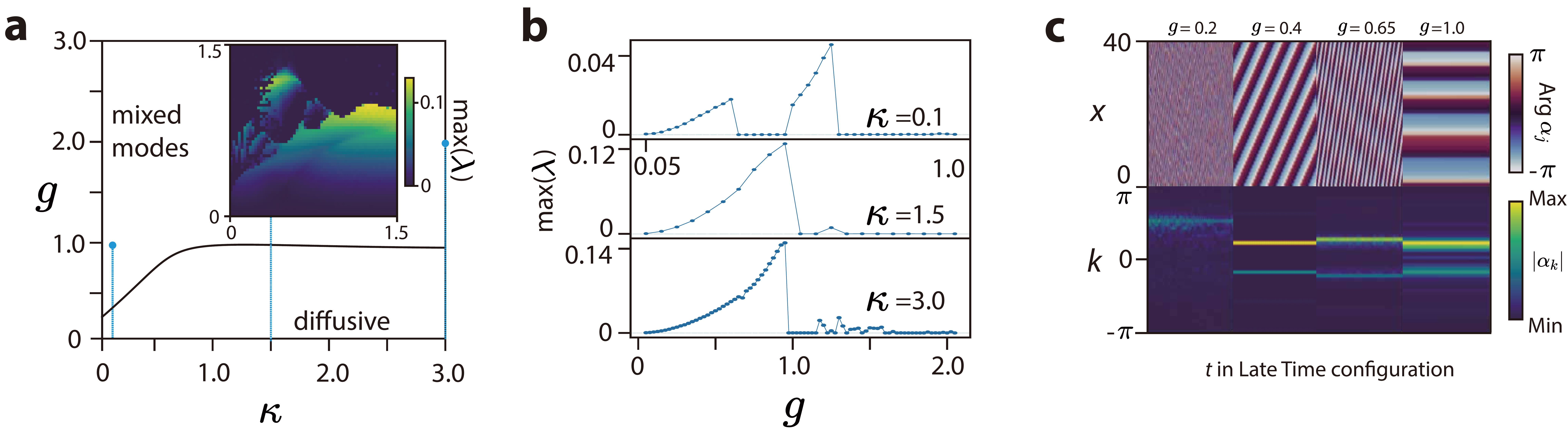}
\caption{\label{fig:SM_PBC}
\textbf{Analysis of mean-field equation of motion under the }periodic boundary condition. \textbf{a.} The sketch of the phase diagram within superradiance of Rabi-Hatano-Nelson model in PBC. Inset: heatmap of the largest Lyapunov exponent of the system. \textbf{b.} Plots of the largest Lyapunov exponent of the system in fixed $\kappa$, corresponding to three blue vertical cuts in (a). \textbf{c.} Long time behavior of the system of $g=0.2$, $0.4$, $0.65$ and $1.0$ from left to right respectively, corresponding to the points sampled on the upper panel ($\kappa=0.1$) in (a). Upper panels plot the phase of the photon fields in real space. Lower panels plot the photon field momentum distributions. Note that the lower panels are not plotted on a same intensity scale. Here, $\omega_0=\Omega=1$, $\theta=\pi$, $N=40$ and $J=0.5$.
}

\end{figure}

Overall, there are two phases, which we call diffusive and mixed-mode phases (Fig.~\ref{fig:SM_PBC}a). In the diffusive phase, the system has a broad momentum distribution (Fig.~\ref{fig:SM_PBC}c leftmost lower panel).
Because light and matter are weakly coupled, a broad range of momenta can be excited and persist for a long time. The photon field behaves like free photons governed by $H_{\text{NH}}$, and the momentum distribution is centered at $k=\pi/2$ (Fig.~\ref{fig:SM_PBC}c leftmost panels), where the lowest loss of the pure photon Hamiltonian $H_{\text{NH}}$ is located.

In contrast, the system has clear peaks of momentum in the mixed-mode phase (Fig.~\ref{fig:SM_PBC}c right lower panels). These peaks exist due to the intrinsic competition between the momentum-asymmetric term of nonreciprocity and the momentum-symmetric term of light-matter coupling through $\Re\alpha_j$.
Furthermore, there are finer structures inside the mixed-mode phase (Fig.~\ref{fig:SM_PBC}a inset and b). When $g$ increases, light–matter coupling suppresses many modes and locks the dynamics to only a few of them. Now momentum asymmetric and symmetric terms lead to the formation of traveling waves (Fig.~\ref{fig:SM_PBC}c second panels from left). As $g$ further increases, these few modes again compete with each other and form few-mode chaos (Fig.~\ref{fig:SM_PBC}c second panels from right). Finally, when $g$ is large enough, light-matter interaction dominates and system evolves into a static state in long time (Fig.~\ref{fig:SM_PBC}c rightmost panels).

\section{Open Boundary Condition}

\subsection{Imaginary Gauge Transformation}
\label{Imaginary Gauge Transformation}

For an open chain, ordinary momentum no longer diagonalizes the nonreciprocal hopping due to the absence of translational symmetry. We first write the full BdG matrix,
\begin{align}
    H_{\text{BdG}} = \begin{pmatrix}
        A & B
        \\
        -B^* & -A^*
    \end{pmatrix}
\end{align}
where the particle block contains the photon Hamiltonian, the local spin energy, and their onsite coupling,
\begin{align}
    A = \begin{pmatrix}
        H_{\text{ph}} & g
        \\
        g & \Omega
    \end{pmatrix}
\end{align}
The effective non-Hermitian photon Hamiltonian carries the directional hopping through its nearest-neighbor matrix elements,
\begin{align}
    (H_{\text{ph}})_{j,j} = \omega_0 -2\iu \kappa,
    \qquad
    (H_{\text{ph}})_{j,j+1} = -J_+,
    \qquad
    (H_{\text{ph}})_{j+1,j} = -J_-,
\end{align}
where $J_+ \equiv -(J+\kappa \e^{-\iu\theta})$ and $J_- \equiv -(J-\kappa \e^{\iu\theta})$.

The anomalous, or pairing, block is purely local:
\begin{align}
    B = \begin{pmatrix}
        0 & g
        \\
        g & 0
    \end{pmatrix}.
\end{align}
Let $r\equiv\sqrt{|J_+/J_-|}$ and introduce the imaginary gauge transformation
\begin{align}
    a_j = r^{j-1} \tilde{a}_j, \qquad b_j = r^{j-1} \tilde{b}_j,
\end{align}
The factor $r^{j-1}$ extracts the exponential envelope associated with the non-Hermitian skin effect \cite{Hatano1996Localization}. In the transformed basis, the hopping amplitudes become reciprocal:
\begin{align}
    -J_+ a^{\dagger}_{j+1} a_j - J_- a_{j} a^{\dagger}_{j+1} = -\sqrt{J_+ J_-} (\tilde{a}^{\dagger}_{j+1} \tilde{a}_j - \tilde{a}_{j} \tilde{a}^{\dagger}_{j+1})
\end{align}
The remaining spatial problem is an ordinary reciprocal chain with open ends. Its normalized standing-wave eigenmodes are
\begin{align}
    \phi_m(j) = \sqrt{\frac{2}{N+1}} \sin(k_m j), \qquad k_m = \frac{m\pi}{N+1}, \quad m=1,\dots,N
\end{align}
with complex eigenvalues
\begin{align}
    \epsilon_m = \omega_0 - 2\iu \kappa - 2\sqrt{J_+ J_-} \cos(k_m).
\end{align}
This simplification does not apply to the entire BdG problem. Particle and hole amplitudes acquire inverse spatial envelopes, so the anomalous terms become strongly position dependent under the same transformation:
\begin{align}
    a_j b_j & = r^{j-1} \tilde{a}_j r^{j-1} \tilde{b}_j = r^{2(j-1)} \tilde{a}_j \tilde{b}_j,
    \\
    a^\dagger_j b^\dagger_j & = r^{-(j-1)} \tilde{a}_j^\dagger r^{-(j-1)} \tilde{b}_j^\dagger = r^{-2(j-1)} \tilde{a}_j^\dagger \tilde{b}_j^\dagger.
\end{align}
Consequently, the standing-wave basis diagonalizes the normal hopping sector but generally mixes different mode indices in the pairing sector. In this basis, the BdG matrix becomes
\begin{align}
    H^\BdG_{mn} = \begin{pmatrix}
        \epsilon_m \delta_{mn} & g \delta_{mn} & 0 & g M_{mn}^{-}
        \\ 
        g \delta_{mn} & \Omega \delta_{mn} & g M_{mn}^{-} & 0
        \\
        0 & -g M_{mn}^{+} & - \epsilon^*_m \delta_{mn} & -g \delta_{mn}
        \\
        -g M_{mn}^{+} & 0 & -g \delta_{mn} & -\Omega \delta_{mn}
    \end{pmatrix},
\end{align}
where
\begin{align}
    M_{mn}^{-} = \sum^N_{j=1} \phi_m(j) r^{-2(j-1)} \phi_n(j),
    \\
    M_{mn}^{+} = \sum^N_{j=1} \phi_m(j) r^{2(j-1)} \phi_n(j)
\end{align}
The matrices $M^\pm$ quantify the mismatch between the particle and hole skin profiles. They are concentrated near the main diagonal when $\kappa<J$ and near the antidiagonal when $\kappa>J$. In the former regime we therefore adopt a diagonal-mode approximation and take $M^\pm$ to commute with $\mathrm{diag}(\epsilon_m)$. This approximation retains the dominant pairing between matching standing-wave modes while neglecting weaker intermode mixing.

An instability occurs when an eigenvalue acquires a positive imaginary part, such that a pair of real eigenvalues coalesces at zero before moving onto the imaginary axis. We proceed by noticing the phase boundary is located by a zero eigenvalue of the mode-resolved BdG matrix. Denoting $\mathcal{E}=\mathrm{diag}(\epsilon_m)$, the zero-eigenvalue condition is
\begin{align}
    0 = \det H_m & = \det\left[\Omega^2 E E^\dagger - g^2 \Omega (E + E^\dagger) \left(1+M^- M^+\right) + g^4 \left(1-M^-M^+\right)^2 \right]
    \\ & = \prod_m\left(\Omega^2 |\epsilon_m|^2 - 2g^2 \Omega (\epsilon_m + \epsilon^*_m)\right)
\end{align}
In the last step, completeness and orthogonality of the standing waves 
remove the overlap matrices within the diagonal-mode approximation. The zero-frequency phase boundary is therefore
\begin{align}
    \left(g_c(\kappa)\right)^2 = \frac{\Omega}{4} \min_m \left\{
        \omega_0 - 2 \sqrt{J_+ J_-} \cos(k_m) + \frac{\kappa^2}{\omega_0 - 2 \sqrt{J_+ J_-}  \cos(k_m)} 
        \right\}, \qquad \kappa<J.
\end{align}
The minimization selects the standing-wave mode that reaches zero first and retains the finite-size quantization through $k_m$. Extending the same reasoning to $\kappa>J$ gives 
\begin{align}
    \left(g_c(\kappa)\right)^2 = \frac{\Omega}{4} \min_m \left\{
        \omega_0 + \frac{1}{\omega_0}\left(\kappa+2 \sqrt{J_+ J_-} \cos(k_m)\right)^2
        \right\}, \qquad \kappa>J.
\end{align}

\subsection{Generalized Bloch Ansatz}

The imaginary-gauge approximation becomes unreliable once intermode pairing is strong. A systematic open-boundary treatment instead uses a generalized Bloch ansatz \cite{Yokomizo2019Non}, in which the spatial factor $\beta$ is allowed to be complex:
\begin{align}
    \Psi_j \propto \beta^j.
\end{align}
Here $|\beta|$ controls the exponential localization of a bulk solution, while its phase controls the spatial oscillation. Substitution into the Schr\"odinger equation gives $H_\BdG(\beta)\Phi=E\Phi$, with
\begin{align}
    H_\BdG (\beta) = \begin{pmatrix}
        \epsilon_p (\beta) & g & 0 & g
        \\
        g & \Omega & g & 0
        \\
        0 & -g & -\Omega & -g
        \\
        -g & 0 & -g & \epsilon_h(\beta)
    \end{pmatrix}.
\end{align}
Unlike the PBC BdG matrix in Eq.~\eqref{eq:PBCBdG}, which restricts $\beta=\e^{\iu q}$ to the unit circle, and the imaginary-gauge BdG approximation, which neglects intermode pairing by treating $M^\pm$ as mode diagonal, $H_{\BdG}(\beta)$ retains both the complex spatial decay and the full particle--hole coupling required under OBC.
The particle and hole dispersions entering this matrix are
\begin{align}
    \epsilon_p (\beta) & = \omega_0 - 2\iu \kappa - J_+ \beta - J_- \beta^{-1},
    \\
    \epsilon_h (\beta) & = -\omega_0 - 2\iu \kappa + J_+^* \beta + J_-^* \beta^{-1}.
\end{align}
Eliminating the internal amplitudes gives the bulk characteristic equation
\begin{align}
    (E^2 - \Omega^2) (E-\epsilon_p(\beta))(E-\epsilon_h(\beta)) - g^2 (\Omega + \epsilon_p (\beta))(\Omega-\epsilon_h(\beta)) = 0.
\end{align}
For $\theta=0$, particle--hole structure allows a more transparent reduction. Defining $\eta(\beta)=\omega_0-J_+\beta-J_-\beta^{-1}$ gives
\begin{align}
    (E^2-\Omega^2)\left((E+2\iu\kappa)^2 - \eta(\beta)^2\right)-g^2\left((\Omega-2\iu\kappa)^2-\eta(\beta)^2\right)=0,
\end{align}
so
\begin{align}
    (\omega_0 -J_+ \beta - J_- \beta^{-1})^2 = \frac{(E^2-\Omega^2)(E+2\iu\kappa)^2 - g^2 (\Omega-2\iu\kappa)^2}{E^2-\Omega^2 -g^2}
    \equiv \Delta(E)
\end{align}
The dependence on $\beta$ can then be factorized into two quadratic polynomials,
\begin{align}
    \left[J_+ \beta^2 - \left(\omega_0 - \sqrt{\Delta(E)} \right) \beta+J_- \right] \left[J_+ \beta^2 - \left(\omega_0 + \sqrt{\Delta(E)} \right)\beta + J_- \right] = 0.
\end{align}
and hence produces four bulk roots for each trial energy:
\begin{align}
    \beta(E) = \frac{\omega + \sigma \sqrt{\Delta(E)} \pm \sqrt{\left(\omega_0 \pm \sqrt{\Delta(E)} \right)^2-4J_+ J_-}}{2J_+}
    \label{eq:OBCroots}
\end{align}
An open-boundary eigenstate is a superposition of these roots. This construction keeps both the skin localization and particle--hole mixing without assuming that the pairing overlaps are diagonal \cite{Yokomizo2021Non}.

\subsection{Spin Effective Equation and Self-partitioned Interfacial Time Crystal}

We next derive a spin-only description that makes the mechanism of the self-partitioned interfacial time crystal (SPITC) more transparent. The starting point is the driven photon equation
\begin{align}
    \partial_t{\alpha}_j = -\iu [H_{\text{HN}} \alpha]_j - \iu g s^x_j
\end{align}
with open-boundary conditions $\alpha_0=\alpha_{N+1}=0$. The photon field is linear in the present approximation and can therefore be integrated out exactly for any prescribed spin trajectory.

Its formal solution separates the freely decaying photon field from the spin-induced response,
\begin{align}
    \alpha(t) = \e^{-\iu H_{\text{HN}} (t-t_0)} \alpha(t_0) - \iu g \int^t_{t_0} \e^{-\iu H_{\text{HN}}(t-t')}s^x(t')\diff t'.
\end{align}
To isolate the slow spin envelope, we move to a frame rotating at the bare precession frequency $\Omega$,
\begin{align}
    S_j^+ = s^x_j + \iu s^y_j = u_j \e^{\iu \Omega t}
\end{align}
so
\begin{align}
    \partial S^+_j = \partial_t u_j \e^{\iu \Omega t} + \iu \Omega u_j \e^{\iu \Omega t} = \partial s^x_j + \iu \partial s^y_j 
    & = -\Omega s^y_j + \iu\Omega s^x_j - 2\iu g (\alpha_j+\alpha^*_j) s^z_j
    \\ & = \iu\Omega u_j \e^{\iu \Omega t} - 2\iu g (\alpha_j+\alpha^*_j) s^z_j .
\end{align}
The $\iu \Omega u_j \e^{\iu \Omega t}$ term on each side of the equation cancels, leaving
\begin{align}
    \partial_t u_j = -2\iu g (\alpha_j+\alpha^*_j) s_j^z \e^{-\iu\Omega t}.
\end{align}
The transverse spin component that drives the photons contains both co-rotating and counter-rotating harmonics. Substituting it into the formal photon solution gives
\begin{align}
    \alpha(t) = \e^{-\iu H_{\text{HN}} (t-t_0)} \alpha(t_0) - \frac{\iu g}{2} \int^t_{t_0} \e^{-\iu H_{\text{HN}}(t-t')} \left( u_j \e^{\iu \Omega t'} + u^*_j \e^{-\iu \Omega t'}\right) \diff t'
\end{align}
We now assume a separation of time scales. The envelope $u$ changes little over the photon relaxation time. At times long compared with that relaxation time, it can be taken outside the integral and the upper integration limit can be extended to infinity,
\begin{align}
    \alpha(t) &= \e^{-\iu H_{\text{HN}} (t-t_0)} \alpha(t_0) - \frac{\iu g}{2} \int^{t-t_0}_{0} \e^{-\iu H_{\text{HN}}\tau} \left( u \e^{\iu \Omega (t-\tau)} + u^* \e^{-\iu \Omega (t-\tau)}\right) \diff \tau
    \\
    & \approx \e^{-\iu H_{\text{HN}} (t-t_0)} \alpha(t_0) - \frac{\iu g}{2} \int^{\infty}_{0} \e^{-\iu  H_{\text{HN}}\tau} \left( u \e^{\iu \Omega (t-\tau)} + u^* \e^{-\iu \Omega (t-\tau)}\right) \diff \tau
    \\
    & = \e^{-\iu H_{\text{HN}} (t-t_0)} \alpha(t_0) - \frac{\iu g}{2} \left( \left(\iu\Omega I + \iu H_{\text{HN}} \right)^{-1} u \e^{\iu \Omega t} + \left(-\iu\Omega I + \iu H_{\text{HN}} \right)^{-1} u^* \e^{-\iu \Omega t}\right).
\end{align}
The first term vanishes at long times because all photon modes are damped in the vacuum phase. The remaining field follows the spin envelope through the photon resolvents,
\begin{align}
    \alpha_j(t) & \approx - \frac{\iu g}{2} \left( [\left(\iu\Omega I + \iu H_{\text{HN}} \right)^{-1} u]_j \e^{\iu \Omega t} + [\left(-\iu\Omega I + \iu H_{\text{HN}} \right)^{-1} u]^*_j \e^{-\iu \Omega t}\right)
    \\
    & = - \frac{\iu g}{2} \left( [R_+ u]_j \e^{\iu \Omega t} + [R_- u]^*_j \e^{-\iu \Omega t} \right),
\end{align}
where $R_\pm\equiv(\pm\iu\Omega I+\iu H_{\text{HN}})^{-1}$. These resolvents are the photon susceptibilities evaluated at the two spin harmonics; their spatial matrix elements describe how a spin fluctuation at one site generates a photon response at another.

Only the real photon quadrature couples back to the spin. It is
\begin{align}
    2\Re\alpha_j = \alpha_j+\alpha_j^* & = \frac{\iu g}{2} \left(
    [\left( R_-^* - R_+ \right) u]_j \e^{\iu \Omega t} - [\left( R_- - R_+^* \right)^* u]_j^* \e^{-\iu \Omega t} \right)
    \\
    & = \frac{\iu g}{2} \left(
    [D u]_j \e^{\iu \Omega t} - [D^* u^*]_j \e^{-\iu \Omega t} \right),
\end{align}
where 
\begin{align}
    D \equiv R_-^* - R_+ .
\end{align}
Inserting this feedback into the transverse-spin equation yields
\begin{align}
    \partial_t u_j & = -2\iu g (\alpha_j+\alpha^*_j) s_j^z \e^{-\iu\Omega t}
    \\
    & = g^2  \left(
    [D u]_j \e^{\iu \Omega t} - [D^* u^*]_j \e^{-\iu \Omega t} \right) s_j^z \e^{-\iu\Omega t}
    \\
    & = g^2  \left(
    [D u]_j - [D^* u^*]_j \e^{-2\iu \Omega t} \right) s_j^z.
\end{align}

The same photon quadrature changes the longitudinal spin component according to
\begin{align}
    \partial_t s^z_j & = 2g (\alpha_j+\alpha^*_j) s^y_j
    = 2g (\alpha_j+\alpha^*_j) \frac{S^+_j - S^-_j}{2\iu} 
    \\ & = -\iu g (\alpha_j+\alpha^*_j) \left( u_j \e^{\iu \Omega t} - u^*_j \e^{-\iu \Omega t} \right)
    \\
    & = \frac{g^2}{2} \left(
    [D u]_j \e^{\iu \Omega t} - [D^* u^*]_j \e^{-\iu \Omega t} \right) \left( u_j \e^{\iu \Omega t} - u^*_j \e^{-\iu \Omega t} \right)
    \\
    & = \frac{g^2}{2} \left( -[D^* u^*]_j u_j - (Du_j) u_j^* + [Du]_j u_j \e^{2\iu \Omega t} + [Du^*]_j u^*_j \e^{-2\iu \Omega t} \right)
    \\
    & = g^2 \left( -\Re\left[[Du]_j u_j^*\right] + \Re\left[ [Du]_j u_j \e^{2\iu \Omega t}\right] \right)
\end{align}

Combining the two results gives a closed, photon-mediated spin dynamics,
\begin{align}
    \partial_t u_j & = g^2  \left(
    [D u]_j - [D^* u^*]_j \e^{-2\iu \Omega t} \right) s_j^z,
    \\
    \partial_t s^z_j & = g^2 \left( -\Re\left[[Du]_j u_j^*\right] + \Re\left[ [Du]_j u_j \e^{2\iu \Omega t}\right] \right).
\end{align}
Terms oscillating at $2\Omega$ average to zero over a complete precession cycle when the envelope varies slowly. After rescaling to the corresponding slow time $T$, the secular dynamics is
\begin{align}
    \frac{\partial}{\partial T} u_j & = s_j^z
    [Du]_j,
    \\
    \frac{\partial}{\partial T} s^z_j & = -\Re\{u_j^*[Du]_j \} .
    \label{eq:SM effective equation}
\end{align}

To evaluate this kernel, it is convenient to separate the bare photon frequency from the dissipative hopping generator by defining
\begin{align}
    -\iu H_\text{HN} = L - \iu \omega_0 I,
\end{align}
such that
\begin{align}
    [L\alpha]_j =-(\Gamma+2\kappa)\alpha_j+\iu(J+\kappa e^{\iu\theta})\alpha_{j+1}+\iu(J-\kappa e^{-\iu\theta})\alpha_{j-1}.
\end{align}
The two resolvents can then be expressed in terms of the sum and difference of the photon and spin frequencies,
\begin{align}
    R_+ & = \left(\iu \Omega I - (L-\iu\omega_0)\right)^{-1} = (\iu \Sigma I - L)^{-1},
    \\
    R_- & = \left(-\iu \Omega I - (L-\iu\omega_0)\right)^{-1} = (\iu \Delta I - L)^{-1},
\end{align}
where we have defined 
\begin{align}
    \Delta = \omega_0 - \Omega, \qquad \Sigma = \omega_0 + \Omega.
\end{align}
This makes the roles of near-resonant and counter-rotating response explicit. The kernel becomes
\begin{align}
    D = (-\iu \Delta I - L^*)^{-1} - (\iu \Sigma I - L)^{-1}
\end{align}
The matrix $L$ can be diagonalized by the same imaginary-gauge construction used in Sec.~\ref{Imaginary Gauge Transformation}. Its open-chain eigenvalues and standing-wave modes are
\begin{align}
    \epsilon^L_m = -2\kappa-\Gamma+2\sqrt{\kappa^2-J^2}\cos(k_m),
\end{align}
\begin{align}
    \phi_m(j) = \sqrt{\frac{2}{N+1}} \sin(k_m j), \qquad k_m = \frac{m\pi}{N+1}, \quad m=1,\dots,N.
\end{align}
The complex-conjugate matrix $L^*$ is related to $L$ by an alternating-sign matrix. We denote it by $\Pi=\mathrm{diag}((-1)^1,(-1)^2,\dots,(-1)^N)$. This operation maps a standing wave at $k_m$ to the reflected wave at $\pi-k_m$, since
\begin{align}
    (-1)^{j+1} \sin(k_mj) = \sin[(\pi-k_m)j] =\sin(k_{\pi-k_m}j),
\end{align}
so
\begin{align}
    \epsilon_m^{L^*} = \epsilon^{L}_{N+1-m} = -2\kappa-\Gamma - \sqrt{\kappa^2 - J^2} \cos(k_m).
\end{align}
Consequently, matrix $D$ is diagonal in the same standing-wave labeling, with eigenvalues
\begin{align}
    d_m = \frac{1}{-\iu \Delta + 2\kappa + \Gamma + \sqrt{\kappa^2 - J^2} \cos(k_m)} - \frac{1}{\iu \Sigma + 2\kappa + \Gamma - \sqrt{\kappa^2 - J^2} \cos(k_m)}.
\end{align}
Near the vacuum, $s_j^z\simeq-1$, so Eq.~\eqref{eq:SM effective equation} linearizes to
\begin{align}
    \partial_T u = -Du,
\end{align}
and the growth rate of each mode is $-\Re[d](k)$. The vacuum loses stability when the largest growth rate reaches zero. At an interior extremum $k^*$, the threshold conditions equivalently are
\begin{align}
    \Re[d](k^*)=0, \qquad \partial_k \Re[d](k^*)=0.
\end{align}

We assume small detuning $\Delta$ now. The additional requirement $|\cos k^*|<1$ ensures that the selected mode lies inside the band rather than at its edge. Solving these conditions perturbatively gives
\begin{align}
    J^2 \leq J^2_\mathrm{c} = \left(\sqrt{2}\Omega - \frac{\Gamma}{2} + \frac{\Delta}{2} - \frac{\Delta^2}{12\sqrt{2}\Omega} \right)^2 - \frac{1}{4} \left(\sqrt{2}\Omega + \frac{\Delta}{\sqrt{2}} - \frac{7\sqrt{2}}{72\Omega} \Delta^2 \right)^2 + \cdots.
\end{align}

In the resonant and lossless-spin limit $\Delta=\Gamma=0$, the critical $J_\mathrm{c}$ reduces to $J_\mathrm{c}=\sqrt{3/2}\Omega$, which corresponds to the case in the main text.
In the following, we discuss the choice of $J$ under two cases, $J<J_\mathrm{c}$ and $J>J_\mathrm{c}$.

\subsubsection{Case $J<J_\mathrm{c}$} 

This situation corresponds to the condition a mode $k^*$ inside the band firstly becomes unstable. In this case, the critical $\kappa_\mathrm{c}$ solves to be
\begin{align}
    \kappa_c = \sqrt{2} \Omega - \Gamma/2 + \frac{\Delta}{2} - \frac{\Delta^2}{12\sqrt{2}\Omega} + O(\Delta^3),
    \label{eq:SM_InternalKappaC}
\end{align}

For clarity, consider the resonant and lossless-spin limit $\Delta=\Gamma=0$. This inequality simplifies to $J_c=\sqrt{3/2}\Omega$. Because $\partial_k\Re[d]$ vanishes at $k^*$, the leading change of the maximal growth rate across the boundary comes directly from varying $\kappa$. At the threshold,
\begin{align}
    \partial_\kappa \Re d = \frac{2}{9\Omega^2},
\end{align}
and therefore the maximal growth rate varies linearly with the distance from criticality:
\begin{align}
    \max\Re[d] \approx \frac{2}{9\Omega^2} (\kappa - \kappa_c). 
\end{align}

\subsubsection{Case $J>J_\mathrm{c}$} 

When $J>\sqrt{3/2}\Omega$, the formal interior extremum leaves the allowed band and the dominant mode is pinned to a band edge. The corresponding critical values are
\begin{align}
    \kappa_{c1} = \sqrt{J^2-\frac{\Omega^2}{4}},
    \quad
    \kappa_{c2} = \frac{2J^2 - \Omega^2}{2\sqrt{J^2-\Omega^2}},
    \label{eq:SM_ExternalKappaC}
\end{align}
The growth rate remains linear in $\kappa-\kappa_c$, although with a different coefficient,
\begin{align}
    \max\Re[d] \approx \frac{2(J^2-\Omega^2)^2}{J^4\Omega^2} (\kappa-\kappa_c)
\end{align}

\begin{figure}
    \centering
    \includegraphics[width=\linewidth]{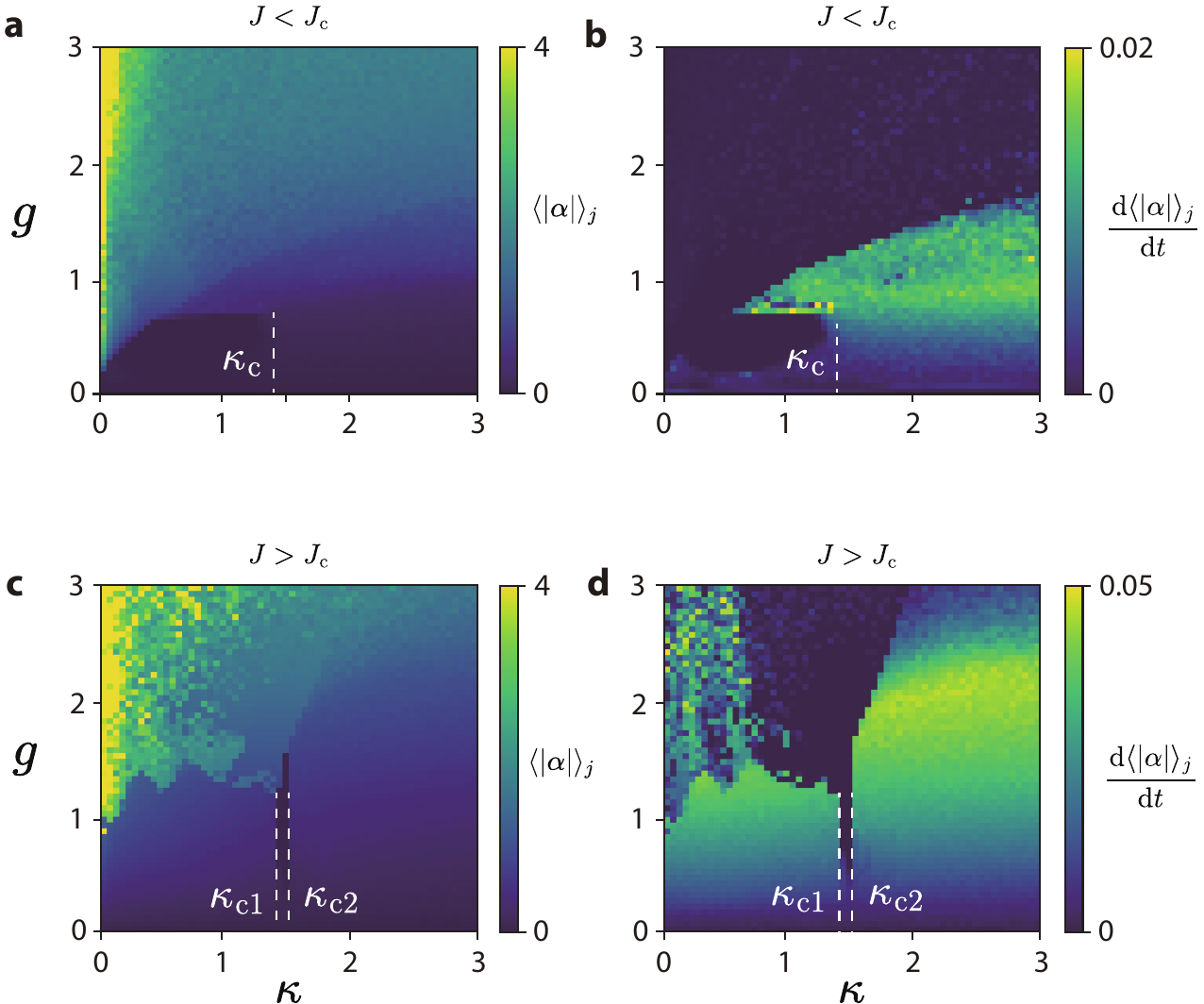}
    \caption{\textbf{Numerical OBC phase diagrams for $J<J_\mathrm{c}$ and $J>J_\mathrm{c}$.} 
    \textbf{a.} $\langle|\alpha|\rangle_j$ when $J=0.5$. \textbf{b.} $\mathrm{d}\langle|\alpha|\rangle_j/\mathrm{d}t$ when $J=0.5$. 
    $\mathbf{a}$ and $\mathbf{b}$ extrapolate to the Figure 2c in the main text.
    \textbf{c.} $\langle|\alpha|\rangle_j$ when $J=1.5$. \textbf{d.} $\mathrm{d}\langle|\alpha|\rangle_j/\mathrm{d}t$ when $J=1.5$. All plots are obtained through numerical integration of Eq.~\eqref{eq:SM_PDE} with parameters $\Omega=\omega_0=1$ and $\Gamma=0$. The white vertical dashed lines in all plots correspond to $\kappa_{\mathrm{c}}$ in Eq.~\eqref{eq:SM_InternalKappaC} and $\kappa_{\mathrm{c1}}$ and $\kappa_{\mathrm{c2}}$ in Eq.~\eqref{eq:SM_ExternalKappaC}.}
    \label{fig:SM_NumPhase}
\end{figure}

\subsubsection{Scalings of the SPITC and the stationary front}

Finally, we relate the spectral structure to the divergence of the interface width $X$. The time-crystal interface, $J/\Omega<\sqrt{3/2}$, is controlled by an in-band momentum, whereas the stationary interface, $J/\Omega>\sqrt{3/2}$, is controlled by a band edge. Expanding $d(k)$ near the relevant $k^*$ gives
\begin{align}
    d(k) = d(k^*) + d'(k^*)(k-k') + \frac{1}{2}d''(k^*)(k-k')^2
\end{align}
For an interior mode, $\partial_k\Re d(k^*)=0$ but the derivative of the imaginary part is generally nonzero. This is the situation at the time-crystal interface,
\begin{align}
    \partial_k d(k^*) = \iu \nu,
\end{align}
where $\nu$ is a nonzero real number. In a finite open region, the momentum can be complex and shift scales inversely with $X$. Matching this shift against the small bulk growth rate gives the marginal-stability condition
\begin{align}
    0 = \Re d(k)= \Re d(k^*) - \frac{\Re[d'(k^*) c]}{X} + O(X^{-2}),
\end{align}
where we used $k \sim c_1/X$ and $c_1$ is a complex number. So
\begin{align}
    X \sim \left(\Re[d](k^*) \right)^{-1} \sim (\kappa-\kappa_c)^{-1}.
\end{align}
At a band edge, by contrast, symmetry removes the entire first derivative,
\begin{align}
    \partial_k d(k^*) = 0.
\end{align}
and the leading finite-$X$ correction is quadratic in the momentum shift. The marginal-stability condition becomes \cite{Belyansky2025Phase}
\begin{align}
    0 = \Re d(k)= \Re d(k^*) - \frac{\Re[d''(k^*) c_2]}{X^2} + O(X^{-3}),
\end{align}
or
\begin{align}
    X \sim \left(\Re[d](k^*) \right)^{-1/2} \sim (\kappa-\kappa_c)^{-1/2}.
\end{align}

\subsection{Numerical Methods}

The red curve in the main text Fig.~4(a) marks the vacuum--condensate interface $X(t)$ extracted automatically from the heat map. For each time slice, we construct an interface score from the positive gradient of the smoothed amplitude profile. The score is weighted to emphasize intermediate amplitudes while suppressing both the vacuum background and the high-amplitude chaotic interior. This makes the detected feature correspond to the transition region rather than to isolated fluctuations on either side.

The resulting score map is analyzed with dynamic programming to identify a globally optimal continuous ridge in space--time. The objective maximizes the cumulative interface score while imposing a quadratic penalty on jumps between adjacent time steps, and a local low-threshold linear interpolation provides sub-grid resolution of $X(t)$. This global continuity constraint prevents brief local defects in the heat map from producing spurious interface jumps.

To determine the intrinsic slow oscillation frequency, we subtract a sliding-time-average background $X_{\text{slow}}(t)$ from the raw trajectory and analyze the detrended signal $\diff X(t)=X(t)-X_{\text{slow}}(t)$. The periodic window is detected automatically from the onset of a finite root-mean-square amplitude together with sufficiently slow variation; the final analysis window begins at the later of this onset and a fixed tail fraction. The dominant frequency is defined as the peak of the power spectrum of $\diff X(t)$ in this window. Separating the slow drift before the spectral analysis avoids misidentifying interface motion or transient relaxation as the time-crystal oscillation.

\bibliographystyle{apsrev4-2}
\bibliography{sources}